\documentclass{webofc}
\usepackage[varg]{txfonts}
\begin{document}
\title{Machine Learning Application for $\mathbf{\Lambda}$ Hyperon Reconstruction in CBM at FAIR}

\author{\firstname{Shahid} \lastname{Khan}\inst{1}\fnsep\thanks{\email{shahidzafarkhan@gmail.com}} \and
        \firstname{Viktor} \lastname{Klochkov}\inst{1}\and
        \firstname{Olha} \lastname{Lavoryk}\inst{2}\and
        \firstname{Oleksii} \lastname{Lubynets}\inst{3,4}\and
        \firstname{Ali Imdad} \lastname{Khan}\inst{1}\and
        \firstname{Andrea} \lastname{Dubla}\inst{3}\and
        \firstname{Ilya} \lastname{Selyuzhenkov}\inst{3,5}}

\institute{
Eberhard Karls University of T{\"u}bingen;
$^2$Taras Shevchenko National University of Kyiv;
$^3$GSI, Darmstadt;
$^4$Goethe University Frankfurt;
$^5$NRNU MEPhI, Moscow
}

\abstract{%
The Compressed Baryonic Matter experiment at FAIR will investigate the QCD phase diagram in the region of high net-baryon densities.
Enhanced production of strange baryons, such as the most abundantly produced $\Lambda$ hyperons, can signal transition to a new phase of the QCD matter.
In this work, the CBM performance for reconstruction of the $\Lambda$ hyperon via its decay to proton and $\pi^{-}$ is presented. 
Decay topology reconstruction is implemented in the Particle-Finder Simple (PFSimple) package with Machine Learning algorithms providing efficient selection of the decays and high signal to background ratio.
}
\maketitle

{\bf Introduction.}
Theoretical calculations predict the possibility of a first order phase transition from hadron gas to a deconfined phase of strongly interacting matter and the existence of a critical point in the region of the QCD phase diagram above 450 MeV chemical potential ($\mu_B$) and below 180 MeV temperature~\cite{li}. 
The experiments at SIS-18, CERN SPS, RHIC beam energy scan and fixed target programs, and future experiments at the Facility for Antiproton and Ion Research (FAIR) and NICA facilities are dedicated to explore this region of the phase diagram. 
The Compressed Baryonic Matter (CBM) experiment at FAIR in Darmstadt will investigate the region at 5--8 normal baryon density ($\mu_B \approx 450-900$~MeV) and temperatures below 120~MeV.
One of the signatures of the new phase of matter is the enhanced production of strange baryons and $\Lambda$ is the most abundantly produced hyperon at FAIR SIS-100 energies.
In this work, the CBM performance for reconstruction of the $\Lambda$ hyperon via its decay to proton and $\pi^{-}$ using Machine Learning algorithms to achieve an efficient selection of the decays and high signal to background ratio is presented.

{\bf CBM experiment setup.}
The CBM experiment consists of three main components: a tracking system inside a dipole magnet, detectors for particle identification and detectors for collision geometry determination.
The tracking system consists of a Micro Vertex Detector (MVD) and a Silicon Tracking System (STS).
The particle identification is provided by the Ring Imaging Cherenkov detector, Muon Chamber, Transition Radiation Detector and Time-of-Flight (TOF) wall. Geometry determination is done with the help of Projectile Spectator Detector, inner part of TOF and the tracking system.
For this analysis a sample of $100k$ minimum bias Au-Au collisions at $12A$ GeV$/c$ was generated using  DCM-QGSM-SMM~\cite{dcm} (Model A) and URQMD~\cite{urqmd} (Model B) with collision products transported through the CBM setup using GEANT4~\cite{geant}.

{\bf Machine learning.}
In the already existing Kalman Filter Particle Finder (KFPF) package~\cite{kfpf} for online reconstruction and selection of short-lived particles in CBM, selection criteria have been manually optimized to maximize signal to background ratio for a certain collision energy and a heavy-ion event generator. 
The selection criteria depend on the collision energy and centrality, decay channel and detector configuration.
Machine Learning (ML) algorithms can be used to adjust these criteria automatically for different data taking scenarios.
ML provides an efficient non-linear and multi-dimensional selection criteria optimization. 
To optimize selection criteria for $\Lambda$ and other particles, ML can use many variables, associated with particle candidates, provided they are not strongly correlated with the invariant mass of the particle.
In this work, we study ML performance using the same variables as used by the KFPF.
Multiple ML libraries were tested for bench-marking using an automated machine learning (automl) package.
The XGBoost~\cite{xgb} model outperformed other models in terms of numerical calculation speed and efficiency. 
XGBoost uses decision trees and a gradient descent optimization algorithm to minimize the loss function.

Details of the $\Lambda$ reconstruction in CBM can be found in~\cite{Particles-2021-4-288}.
Cellular automaton and Kalman Filter Particle (KFParticle) based PFSimple package~\cite{Particles-2021-4-288} are used for track finding, fitting and decay kinematics reconstruction.
PFSimple interfaces the mathematics of the KFParticle package and provides a convenient interface to control the reconstruction parameters. 
The $\Lambda$ candidates constructed from proton and $\pi^-$ track pairs coming from a true $\Lambda$ decay are termed as signal while all other candidates are considered as background.

The list of variables used for ML includes a squared distance between the daughter tracks and primary vertex (PV) divided by its error, the closest distance between the tracks (DCA), a squared distance between daughter tracks divided by its error, and distance between PV and secondary vertex divided by its error.
A set of preselection criteria are applied to the data to remove numerical artefacts. 
A central Au-Au collision contains mostly combinatorial background and way less signal for $\Lambda$. 
Since the data has imbalanced classes, the data set used to train and test the ML algorithm is over-sampled by increasing the ratio of signal (under-represented class) to background candidates.
Model A is treated as simulated signal and used for efficiency calculation while Model B is treated as background (which in real data analysis is taken from experimental data).

\begin{figure*}
\centering
\includegraphics[width=0.5\textwidth]{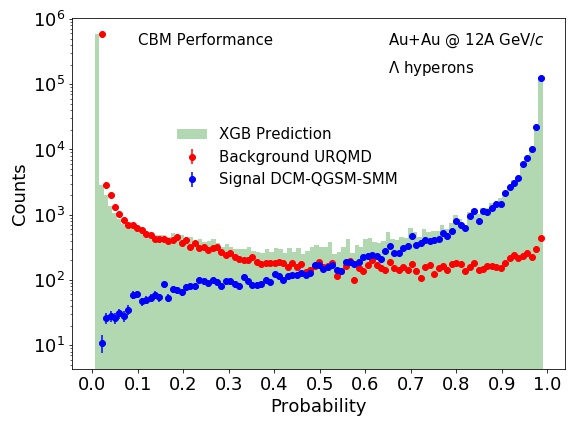}%
\includegraphics[width=0.5\textwidth]{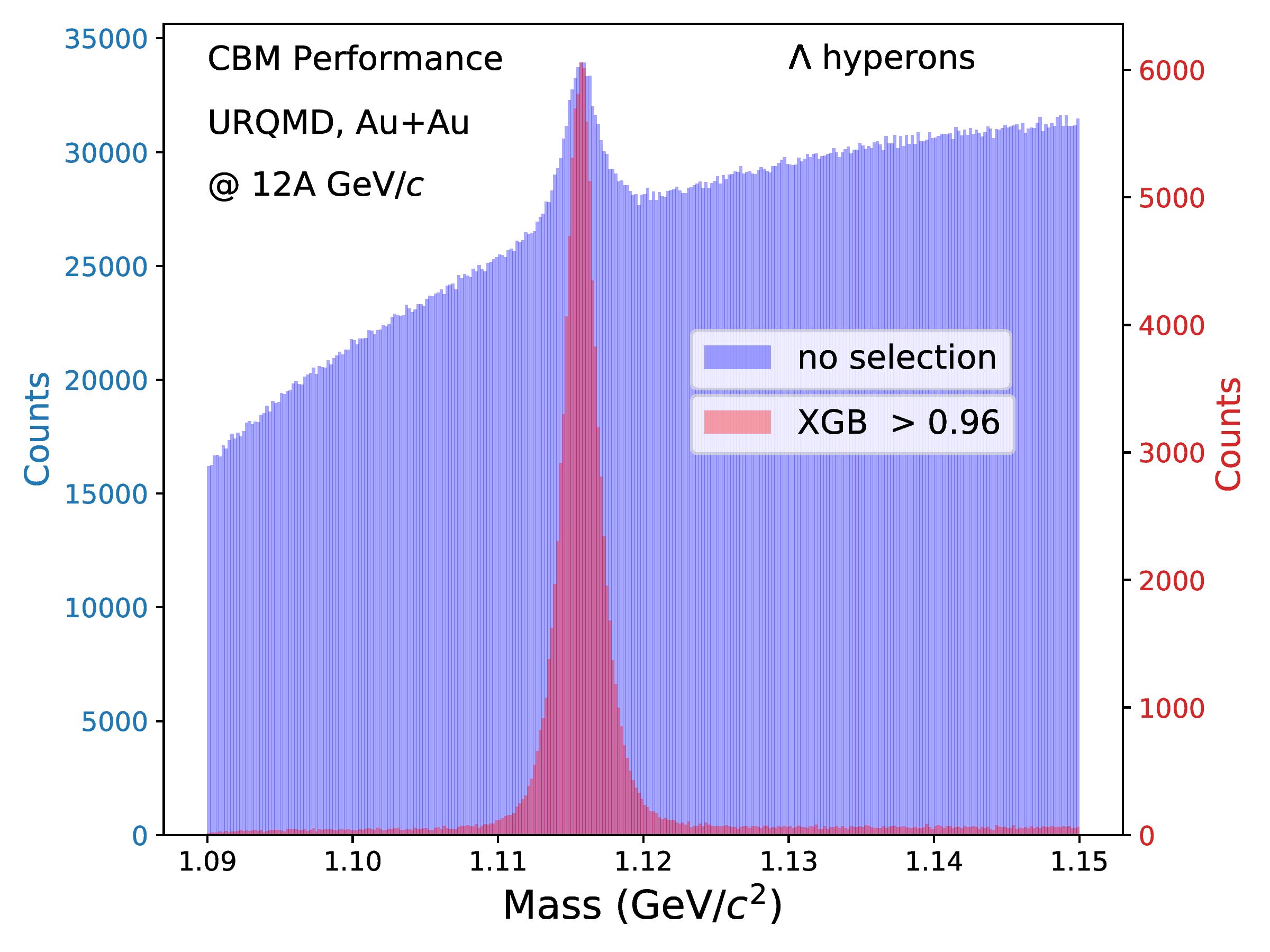}
\caption{
(left) Performance of the XGB algorithm for signal--background separation in the $\Lambda$ candidates.
The arrow represents the maximum AMS value at which selection was applied.
(right) The invariant mass distribution of $\Lambda$ candidates before (blue) and after XGB selection (red).}
\label{prob-cut}
\end{figure*}

A sample of about 1M of $\Lambda$ signal from Model A in the $5\sigma$ region around the $\Lambda$ peak in the invariant mass distribution and a sample of about 3M of the background from Model B on the left and right hand sides outside of the $\Lambda$ peak region and below 1.3 Gev/$c^2$ were used for the analysis. 
Using background from a different model (Model B) helps to reduce over fitting and dependence of the ML result on the signal model. 
These samples are divided in proportion 80\% to 20\% into train-test samples. 
To tune various parameters of the algorithm so that it fits better on the training data and performs well on the test data, Bayesian optimisation~\cite{bayesian} package is used. 
It divides the train data into 5 folds and searches for optimal values of parameters using parallel computing. 
Area Under the Curve (AUC) parameter is used to select the best model. 
The trained model is then applied on the test data set and it returns a probability distribution between 0 and 1 for given input variables as shown in Figure~\ref{prob-cut} (left). 
Candidates which are close to 0 are most probably background and vice versa for candidates close to 1. 
Based on this probability distribution an approximate median significance (AMS)~\cite{ams} is calculated to maximise signal to background ratio.

{\bf Results and discussion.}
The trained and tested XGB model, along with the AMS selection criteria, is then applied on two samples for both Models A and B containing $100k$ events each.
Figure~\ref{prob-cut} (right) shows $\Lambda$ candidates from Model B (UrQMD) before and after XGB selection is applied. 
For differential analysis the kinematic phase space is divided in transverse momentum ($p_{\mathrm{T}}$) intervals of 0.5 GeV/c wide and laboratory rapidity ($y_{\mathrm{lab}}$) intervals of 0.5 step size.
Efficiency of the ML selection is calculated by dividing the number of remaining true signal (true positives in the confusion matrix) in the XGB selected $\Lambda$ candidates by the number of input signal for each $p_{\mathrm{T}}$-$y_{\mathrm{lab}}$ interval. 
\begin{figure*}
\centering
\includegraphics[width=0.42\textwidth]{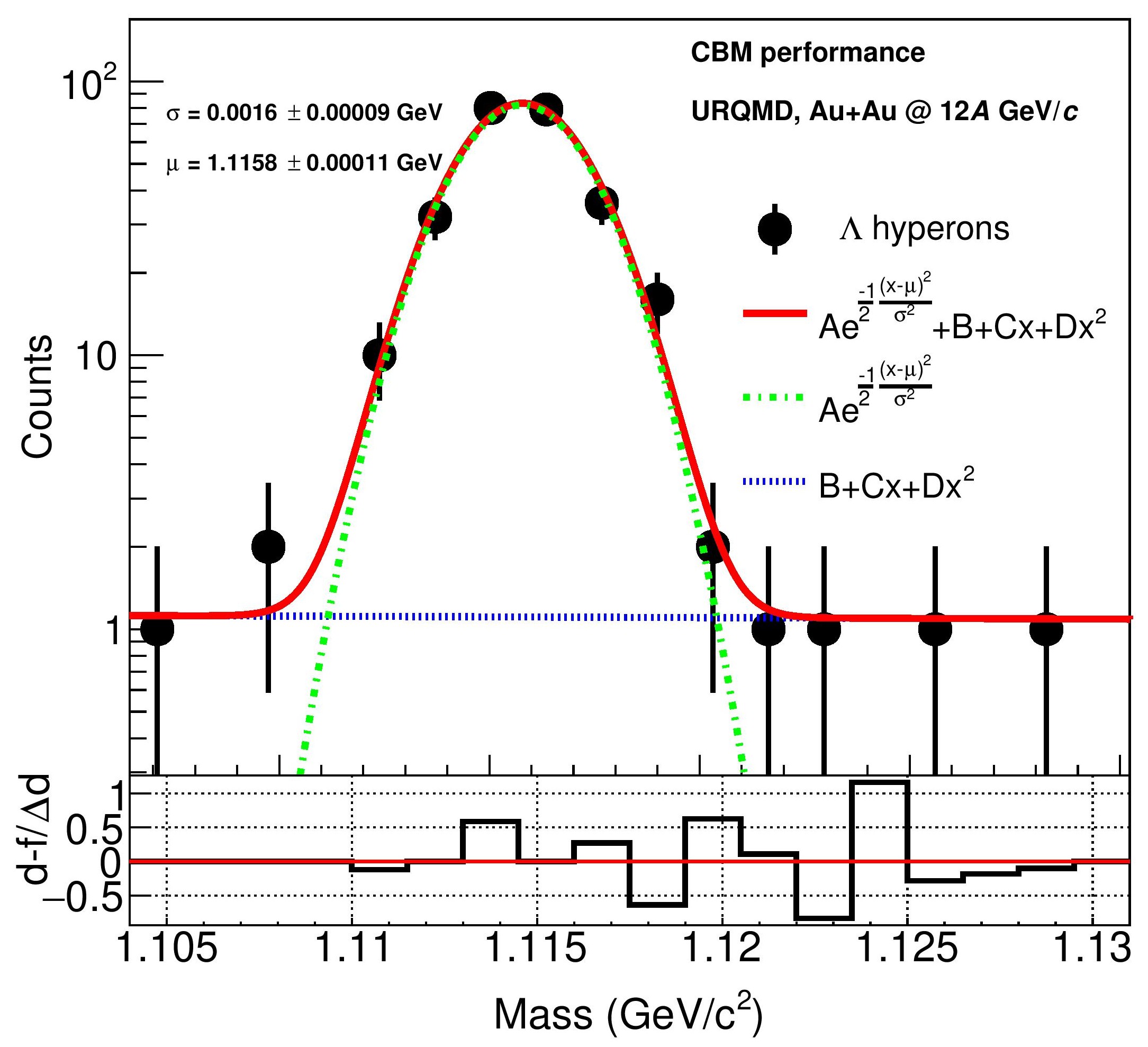}~~%
\includegraphics[width=0.56\textwidth]{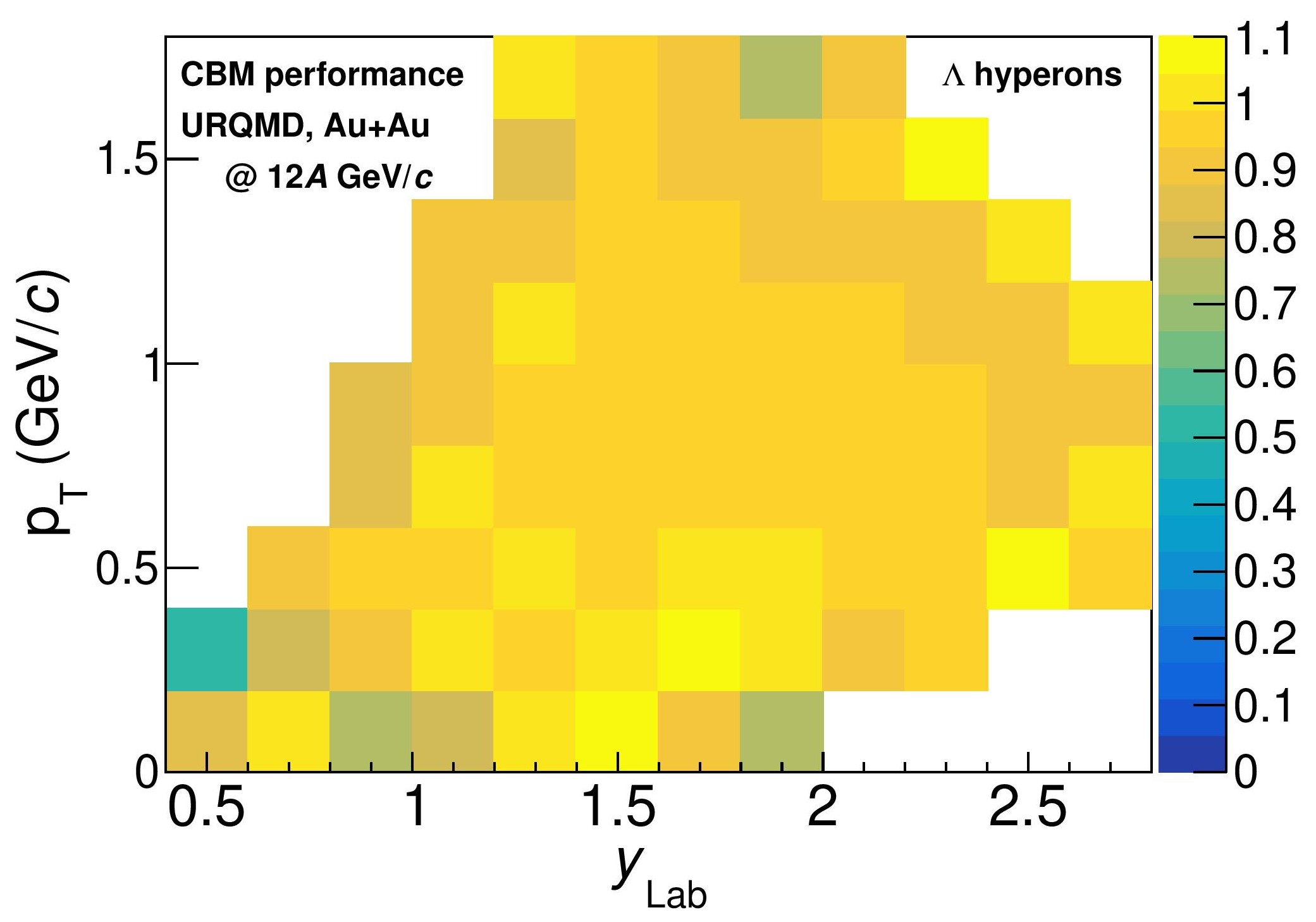}
\caption{
(left) Illustration of the procedure for $\Lambda$ signal extraction after XGB selection is applied.
Example is for $1.4 < p_{\mathrm{T}} < 1.6$~GeV/$c$ and $2.2 < y_{\mathrm{lab}} < 2.4$ interval.
The invariant mass yield (black circles) is fitted with the signal (Gaussian, green dotted line) and background (second order polynomial, blue line) shapes.
(right) The $p_{\mathrm{T}}$-$y_{\mathrm{lab}}$ distribution for the ratio of the fully corrected yield extracted after XGB selection to the simulated input.
}
\label{fit-2dhist}
\end{figure*}

\begin{figure}[h]
\centering
\includegraphics[width=0.8\textwidth]{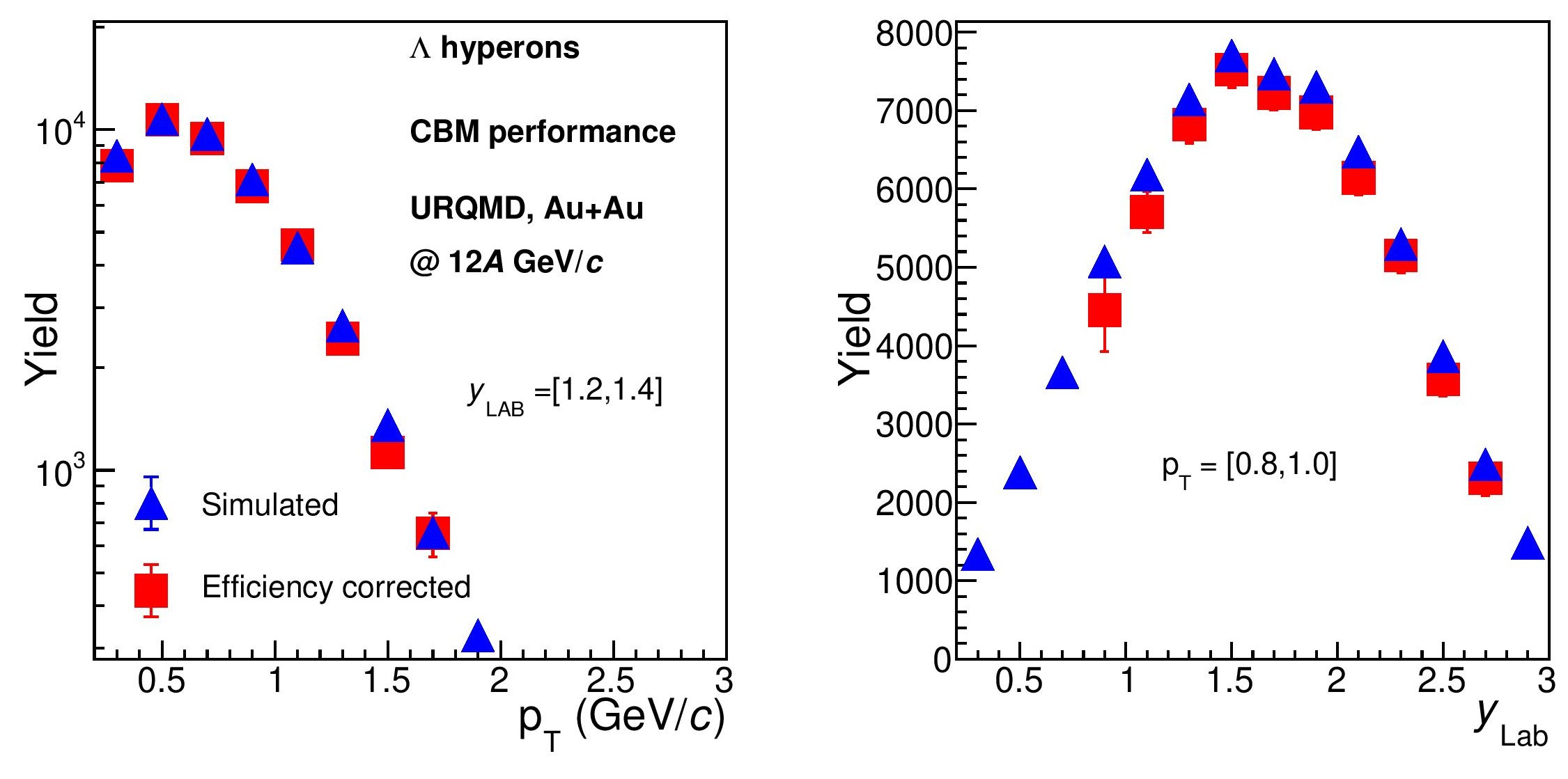}
\caption{
Yield of $\Lambda$ hyperons as a function of (left) transverse momentum, $p_{\rm T}$, for rapidity $1.2 < y_{\mathrm{lab}} < 1.4$ and (right) rapidity for $0.8 < p_{\mathrm{T}} < 1.0$.
}
\label{yield}
\end{figure}

The yield extraction is implemented as a three stage fitting procedure on Model B to each $p_{\mathrm{T}}$-$y_{\mathrm{lab}}$ interval. 
Only the ranges which contain more than 400 candidates are considered.
In stage one, the invariant mass distribution of the background, $\Lambda$ candidates outside the $5\sigma$ region from the $\Lambda$ peak at 1.115~GeV/$c^2$, is fitted with a second order polynomial ($pol2$).
In the second stage, the invariant mass distribution is fitted in full range with a sum of a Gaussian ($A \exp[((x-\mu)/\sigma)^2/2]$) and $pol2$ functions. 
For this fit the mean ($\mu$) and standard deviation ($\sigma$) of the Gaussian function are fixed to $\mu = 1.115$ GeV/$c^2$ and $\sigma = 0.0014$ GeV/$c^2$, while the initial values of the $pol2$ are taken from the fit at stage one. 
In the final stage, all parameters of the Gaussian and pol2 are released with their initial values set to the result of the fit at stage two. 
The number of entries under the Gaussian fit only, in the $2.5\sigma$ region around $\mu$, are classified as $\Lambda$ signal. 
Figure~\ref{fit-2dhist} (left) shows an example of the final result from the fitting procedure.
The corrected $\Lambda$ yield for each $p_{\mathrm{T}}$-$y_{\mathrm{lab}}$ interval is extracted by dividing the signal yield from the fitting procedure with the efficiency correction factor obtained from Model A. 
The corrected yield divided by the simulated yield is plotted in Figure~\ref{fit-2dhist} (right).
Figure~\ref{yield} demonstrates good agreement of the corrected yield with the simulated one as a function of $p_{\mathrm{T}}$ (left) and $y_{\mathrm{lab}}$ (right).

{\bf Summary.}
The CBM performance for reconstruction of the $\Lambda$ hyperon via its decay to proton and $\pi^{-}$ in Au-Au collisions for the top FAIR SIS-100 beam momentum is presented.
By combining the precise Kalman-Filter based reconstruction of the decay topology and machine learning algorithms a multi-differential analysis of the $\Lambda$ yield in $p_{\mathrm{T}}$ and $y_{\mathrm{lab}}$ was performed.
Reliable extraction of the $\Lambda$ hyperon is demonstrated by comparison of the efficiency corrected yields after the reconstruction and ML selection with the simulated input for different heavy-ion event generators DCM-QGSM-SMM and UrQMD.
In future, the method will be deployed for different collision energies and other strange and multi-strange particles and (hyper-)nuclei decays.

{\bf Acknowledgements.}
This work was supported by the Carlo and Karin Giersch Stiftung, HGS-HIRe Graduate School by HIC for FAIR, the Ministry of Science and Higher Education of the Russian Federation, Project “Fundamental properties of elementary particles and cosmology” No 0723-2020-0041, the Russian Foundation for Basic Research (RFBR) funding within the research project no. 18-02-40086, and the European Union’s Horizon 2020 research and innovation program under grant agreement No. 871072.

\end{document}